\documentstyle[aps,prl,multicol,epsfig]{revtex}

\begin{document}
\draft
\title{Fano and Kondo resonance in electronic current through nanodevices}
\author{Bogdan R. Bu{\l}ka and Piotr Stefa\'nski}

\address
{Institute of Molecular Physics, Polish Academy of
Sciences, ul.Smoluchowskiego 17, 60-179 Pozna\'n, Poland}
\date{Received \hspace{5mm} }
\date{\today}
\maketitle

 \thispagestyle{empty}
\begin{abstract}

 Electronic transport through a quantum dot strongly coupled to electrodes
is studied within a model with two conduction channels. It is
shown that multiple scattering and interference of transmitted
waves through both channels lead to Fano resonance associated with
Kondo resonance. Interference effects are also pronouncedly seen
in transport through the Aharonov-Bohm ring with the Kondo dot,
where the current characteristics continuously evolve with the
magnetic flux.
 \end{abstract}
\pacs{PACS 73.63.-b, 72.15.Qm, 75.20.Hr} \vskip -1cm
\begin{multicols}{2} \narrowtext
Recent electron transport experiments performed in a single
electron transistor strongly coupled to electrodes \cite{gores}
and by a scanning tunneling microscope (STM) on a single magnetic
adatom on a metallic surface \cite{schneider,madhavan} showed that
the Kondo resonance \cite{hewson} occurs simultaneously with the
Fano resonance \cite{fano}. Multiple scatterings of travelling
electronic waves on a localized magnetic state are crucial for a
formation of both resonances. The condition for the Fano resonance
to appear is a presence of at least two scattering channels: the
discrete level and the broad continuum band \cite{fano,fanot}. In
the mesoscopic systems the nature of two conduction channels is
dependent on the geometry of the device under consideration.
Interferometer geometry is realized when an Aharonov-Bohm ring
with a quantum dot (QD) placed in one of the arms is studied
\cite{yacoby,gerland,wiel}. When an adatom is deposited on the
metallic surface, the STM tip probes indirectly the hybridized
local adatom level together with the band of surface electrons
\cite{schneider,madhavan,schiller,ujsaghy}. We consider the
transmission geometry, when the coupling of the QD to the leads
increases to a strong regime \cite{gores} and additional
transmission channels are activated. The QD is a multilevel system
and the transmission through a higher level (close to the Fermi
energy) can be treated as an effective bridge channel. Although
the electron transport through the QD is governed by the Kondo
effect, interference processes are essential and can produce the
Fano-shaped resonances. It is also interesting to analyze the
Aharonov-Bohm ring with the Kondo impurity. In such a system one
can continuously change the interference conditions by varying a
magnetic flux and can observe the resulting evolution of the
current characteristics from the Kondo peak to the Fano dip. The
studies have been performed for various energies of the impurity
state: in the Kondo regime, in the mixed-valence regime as well as
in the empty state regime.

Our model is described by the Hamiltonian

\begin{eqnarray}\label{1}
H = \sum_{k,\sigma, \alpha\in L,R} \epsilon_{k\alpha}\;
c_{k\alpha,\sigma}^{\dagger}c_{k\alpha,\sigma}
\nonumber\\+\sum_{\sigma}[\epsilon_0
\;c_{0\sigma}^{\dagger}c_{0\sigma} +
\frac{U}{2}\;c_{0\sigma}^{\dagger}c_{0\sigma}c_{0-\sigma}^{\dagger}c_{0-\sigma}
]\nonumber\\ +\sum_{k,\sigma}
[t_{L0}\;c_{kL,\sigma}^{\dagger}c_{0\sigma}+t_{R0}\;
c_{kR,\sigma}^{\dagger}c_{0\sigma}+h.c.]\nonumber\\
+\sum_{k,k',\sigma}
[t_{LR}\;c_{kL,\sigma}^{\dagger}c_{k'R,\sigma}+h.c.]\;.
\end{eqnarray}

The first term describes electrons in the in the left (L) and the
right (R) electrode; the second one describes the quantum dot with
a single state $\epsilon_0$ and Coulomb interactions characterized
by the parameter $U$; the third one corresponds to the tunneling
from the
 electrodes to the dot; and the last one
describes the {\it bridge} channel over the dot.

The current from the left electrode can be calculated from the
time evolution of the occupation number
$N_L=\sum_{k,\sigma}c_{kL,\sigma}^{\dagger}c_{kL,\sigma}$ for
electrons in the left electrode using the Green functions of the
Keldysh type~\cite{meir}. The result is
\begin{eqnarray}\label{3}
J =\frac{2e}{\hbar} \int \frac{d\omega}{2\pi}{\mathrm{ Re}} \Bigl[
\sum_{k,\sigma} t_{L0}\; G_{0,kL\sigma}^{<}(\omega)\nonumber\\+
\sum_{k,k',\sigma}
t_{LR}\;G_{k'R\sigma,kL\sigma}^{<}(\omega)\Bigr]\;,
\end{eqnarray}
where $G_{0,kL\sigma}^{<}(\omega)$ and
$G_{k'R\sigma,kL\sigma}^{<}(\omega)$ are the lesser Green
functions corresponding to the states at the dot and in the left
electrode or the states in both the electrodes, respectively.
Next, we use the Dyson equation to calculate the non-equilibrium
Green functions and express $J$ only by the Green function
$G_{00}$ at the dot and the bare Green functions $g_{\alpha}$ in
the electrodes. The lesser, retarded and advanced Green functions
$g_{\alpha}$ are taken in the form $g^<_{\alpha}=2i\pi \rho
f_{\alpha}$ and $g^{r,a}_{\alpha}=\mp i\pi \rho$, where
$f_{\alpha}$ denotes the Fermi distribution function for electrons
in the $\alpha$-electrode and $\rho$ is the density of states. All
multiple scatterings on the dot and the contacts, as well as
interference processes, are taken into account. Assuming
quasi-elastic transport, for which the current conservation rule
is fulfilled for any energy $\omega$, one obtains
\begin{eqnarray}\label{5}
J =\frac{2e}{h} \int
d\omega[f_L(\omega)-f_R(\omega)]\Bigl\{\alpha_{LR}\;|t_{LR}|^2
\nonumber\\+{\mathrm{ Im}}[\alpha_{00}]\;{\mathrm{ Re}}
[G^r_{00}(\omega)]+{\mathrm{ Re}}[\alpha_{00}]\;{\mathrm{ Im}}[
G^r_{00}(\omega)]\Bigr\}\;,
\end{eqnarray}
where $\alpha_{LR}=4\pi^2\rho^2/w^2$, $\alpha_{00}=-4\pi\rho
z_L^-z_L^{+*}z_R^-z_R^{+*}/$ $[w^2(|z_{L}|^2+|z_{R}|^2)]$,
$w=1+\pi^2\rho^2|t_{LR}|^2$, $z_L^{\pm}=t_{L0} \pm i\pi\rho
t_{LR}t_{R0}^*$, $z_R^{\pm}=t_{R0}^* \pm i \pi\rho
t_{LR}^*t_{L0}$. The formula (\ref{5}) includes two new terms: the
first one corresponding to the current through the bridge channel
and the second one proportional to Re$G^r_{00}(\omega)$, which is
responsible for the Fano resonance~\cite{fano}.

In order to determine the Green function $G^{r}_{00}$ we choose,
among a few known approaches~\cite{hewson}, the equation of motion
(EOM) method. Although the EOM describes the Kondo resonance only
qualitatively, it takes into account all relevant interference
processes and can be applied straightforwardly for our model
(\ref{1}) with the bridge channel. The method generates
higher-order Green functions, which are truncated according  to
the self-consistent decoupling procedure proposed by
Lacroix~\cite{lacroix}. In the limit $U\to \infty$ the Green
function at the dot is determined as
\begin{eqnarray}\label{6}
G^r_{00}(\omega)=\frac{1-n/2-a_{00}}{\omega-\epsilon_0
+i\Delta_0-2i\Delta_0 a_{00}-b_{00}}\;,
\end{eqnarray}
where $\Delta_0=\pi\rho(|t_{L0}|^2+|t_{R0}|^2)$,
$a_{00}=(|z_L|^2H_L(\omega)+|z_R|^2H_R(\omega))/w$ and
$b_{00}=(|z_L|^2F_L(\omega)+|z_R|^2F_R(\omega))/w$.
 Here, we use the functions
\begin{eqnarray}\label{10}
H_{\alpha}(\omega)=\frac{\rho}{w}\int_{-D}^{D}
\frac{d\omega'f_{\alpha}(\omega')[G^r_{00}(\omega')]^*}{\omega-\omega'}\;,\\
\label{10a}F_{\alpha}(\omega)=\frac{\rho}{w}\int_{-D}^{D}
\frac{d\omega'f_{\alpha}(\omega')}{\omega-\omega'}=\nonumber\\
\frac{\rho}{w}\Bigl\{i\pi f_{\alpha}(\omega)+\ln\frac{2\pi
k_BT}{D}+{\mathrm{
Re}}\Psi\Bigl[\frac{1}{2}-i\frac{\omega-\epsilon_{F\alpha}}{2\pi
k_BT}\Bigr]\Bigr\}\;,
\end{eqnarray}
where $\Psi$ is the digamma function and $\epsilon_{F\alpha}$
denotes the position of the Fermi level in the $\alpha$ electrode.
Derivations were performed for the constant density of states
$\rho (\epsilon)=1/2D$ for $|\epsilon| <D$ (in our further
calculations $D$ is taken as unity). The Green function (\ref{6})
is similar to that one obtained by Lacroix~\cite{lacroix} if one
exchanges the tunneling matrix $t_{\alpha0}$ by an effective one
$z^r_{\alpha}$. The electron concentration at the dot is given by
\begin{eqnarray}\label{11}
n=-\frac{2}{\pi} \int d\omega f_0(\omega){\mathrm{
Im}}[G^r_{00}(\omega)]\;,
\end{eqnarray}
where $f_0(\omega)=\gamma_Lf_L(\omega)+\gamma_Rf_R(\omega)$ is the
non-equilibrium distribution function at the dot,
$\gamma_L=|z_L|^2/(|z_L|^2+|z_R|^2)$ and
$\gamma_R=|z_R|^2/(|z_L|^2+|z_R|^2)$.

At $T=0$ the functions $H_{\alpha}(\omega)$ and
$F_{\alpha}(\omega)$ have logarithmic singularities at
$\omega=\epsilon_{F\alpha}$, but $G_{00}(\omega)$ varies more
smoothly around this point. At equilibrium the equation (\ref{6})
can be written as
\begin{equation}\label{11a}
G^r_{00}(\epsilon_F)=\frac{[G^r_{00}(\epsilon_F)]^*}{2i\Delta_0
[G^r_{00}(\epsilon_F)]^*+1}\;.
\end{equation}
A solution of this equation is
$G^r_{00}(\epsilon_F)=[1-e^{2i\phi}]/(2i\Delta_0)$, where the
phase $\phi$ is taken according to the Friedel sum
rule~\cite{hewson} as $\phi=\pi n/2$ . On insertion of $G^r_{00}$
into (\ref{5}) one can find the  conductance (for the zero bias
voltage $V\to 0$)
\begin{eqnarray}\label{12}
{\cal
G}=\frac{2e^2}{h}\bigl\{\alpha_{LR}|t_{LR}|^2-\frac{1}{2\Delta_0}{\mathrm{
Re}}[\alpha_{00}(1-e^{i\pi n})]\bigr\}\;.
\end{eqnarray}
The electron concentration $n$ can be associated with the relative
position of the level $\Delta\epsilon=\epsilon_F-\epsilon_0$
as~\cite{hewson} $n=\frac{1}{2}+ \frac{1}{\pi}
\tan^{-1}[\Delta\epsilon/\Delta_0]$, which can be varied by the
gate voltage applied to the dot.

 For finite temperatures the  set of the
self-consistent equations (\ref{6}), (\ref{10}) and (\ref{11}) is
solved numerically. Figure 1a presents the results for the
zero-bias conductance ${\cal G}$ through the system with the Kondo
dot only ($t_{LR}=0$). The Kondo temperature is within the EOM
\cite{lacroix} $T_K=0.57\exp(-\pi\Delta\epsilon/\Delta_0)/k_B\rho$
and the cross-over from the mixed-valence to the Kondo regime is
at
$\Delta\epsilon_{cr}=\Delta_0/\pi-\Delta_0/\pi\ln(\rho\Delta_0/2\pi)$,
which gives $\Delta\epsilon_{cr}\approx 0.056$ and
${\mathrm{max}}(T_K)\approx 0.004$ for the parameters used in
Fig.1. At $T=0$, ${\cal G}$ is a step-like function, whose maximum
value \vskip 0.1cm\begin{figure}[h]
\centerline{\epsfxsize=0.34\textwidth \epsfbox{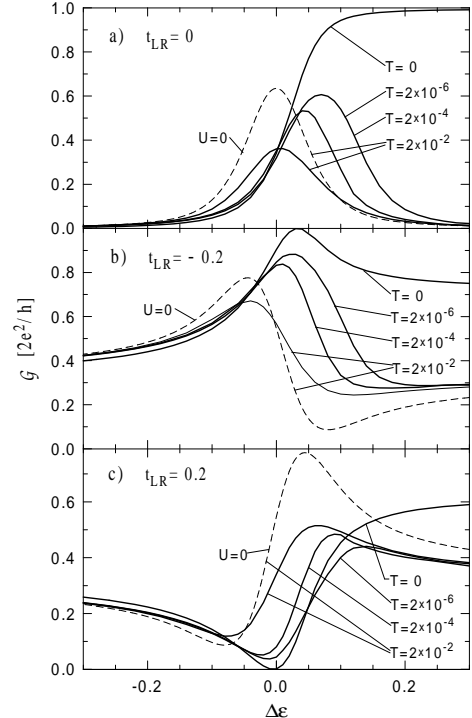} }\vskip
0.1cm\caption{Conductance through the Kondo system with $t_{LR} =
0$ (Fig.a), $t_{LR} = -0.2$ (Fig.b) and $t_{LR} = 0.2$ (Fig.c) as
a function of $\Delta\epsilon$ for $T= 0$, $2\times 10^{-6}$,
$2\times 10^{-4}$ and $2\times 10^{-2}$. For comparison ${\cal G}$
is shown for the dot without Coulomb interactions ($U=0$) at a
high temperature $T=0.02$ (the dashed curves). In all our
calculations $t_{L0}=0.1$, $t_{R0}=0.1$, for which
$\Delta_0=0.031$. }\end{figure} $2e^2/h$ is reached in the Kondo
regime (see also ~\cite{kondo}). At very high temperature $T\gg
T_K$ Coulomb interactions are irrelevant, and the conductance peak
is at $\Delta\epsilon=0$ (see the curve corresponding to $T=0.02$
in Fig.1a). When $T<T_K$ the peak is shifted to the Kondo regime
and increases logarithmically for $T\to 0$. Such a shift was
observed in many experiments on quantum dots~\cite{kondoexp}.
Since the EOM underestimates temperature dependences, one can
expect more pronounced temperature changes for ${\cal G}$ than
predicted by this method.

Figures 1b and 1c present the results for the system with the
bridge channel. The conductance curves have an asymmetric shape,
which is typical for a Fano resonance. The effect can be clearly
visible for  $|t_{LR}|\gtrsim |t_{L0}|, |t_{R0}|$. If $t_{LR}$ is
negative (Fig.1b), ${\cal G}$ exhibits a large maximum, whereas a
deep minimum exists for positive values of $t_{LR}$ (Fig.1c). This
results from constructive and destructive interference processes
for electrons transmitted through two channels. Comparing ${\cal
G}$ for the interacting and the noninteracting case (the solid and
the dashed curves, respectively, in Fig.1) one sees that
correlations on the dot weaken the Fano resonance effect (the
maximum and the minimum for the solid curves are smaller in the
Kondo regime than those for the dashed curves).

In a similar way we calculated the source-drain voltage
characteristics of the device. It was assumed that the potential
$V$ is applied to the left electrode and in the right electrode
the potential is kept zero. Figure 2 presents the evolution of the
differential conductance $dI/dV$ with the variation of the
relative position of the impurity level $\Delta\epsilon$, from the
Kondo to the mixed-valence regime. The case for the pure Kondo dot
($t_{LR}=0$) is given in Fig.2a. If the impurity level lies in the
Kondo regime, the curves show a very narrow peak at low voltage
due to the Anderson hybridization. The sharp feature in the
mixed-valence regime (the dash-dotted curve in Fig.2a) we
attribute to the EOM method. The peak disappears when
$\Delta\epsilon$ approaches the empty state regime (see the dotted
curve). Out-equilibrium the Kondo peak is split into two peaks,
which are pinned to $\epsilon_F+eV$ and $\epsilon_F$ (i.e. to the
chemical potentials
 of the left and the right electrode,
respectively). This is manifested in the narrow peak of $dI/dV$,
which was observed experimentally as well~\cite{kondoexp}. The
broad maximum seen in Fig.2a results from the resonant tunneling
when the chemical potential $\epsilon_F+eV$ approaches
$\epsilon_0$.

 The influence of the
bridge channel is presented in Figs.2b and 2c. A direct electron
transmission increases the differential conductance (for our case
the bridge channel contribution to ${\cal G}$ is $0.33\times
2e^2/h$). The $dI/dV$ curves in Fig.2b are similar to those in
Fig.2a. Although the broad resonance maxima are deformed, there
are well pronounced narrow peaks in the low voltage regime. The
situation for $t_{LR}= 0.2$, presented in Fig.2c, is, however,
different. In the low voltage range the curves show a narrow dip
instead of a peak. In this case there is a
\begin{figure}[h]\label{figfk2}
\centerline{\epsfxsize=0.37\textwidth
\epsfbox{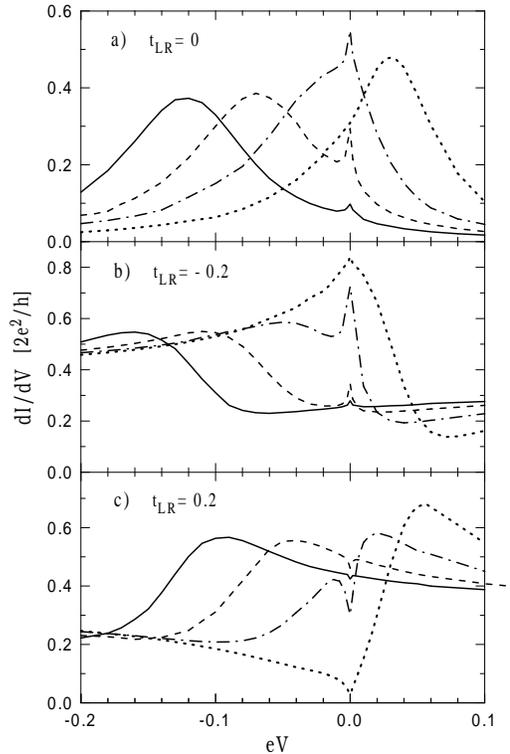}}\vskip0.1cm
 \caption{Differential conductance
as a function of the applied voltage for $\Delta\epsilon= 0.15$
(solid curve), 0.1 (dashed curve), 0.05 (dash-dotted curve) and 0
(dotted curve) at $T=2\times 10^{-4}$.}
\end{figure} destructive
interference of electronic waves passing through the bridge and
the Kondo dot. If a small voltage is applied the transmission
through the Kondo dot is lowered, which weakens interference of
the two channels. It results in the opening of the bridge channel
and increase of the conductance. We found that temperature and
voltage characteristics for the dip are similar to those for the
low voltage peak in the Kondo dot. It is not surprising as the
Kondo resonance plays a crucial role in both the situations.

 Recent
experiments performed by G{\"o}res et al. \cite{gores} showed that
the gray-scale plot of $dI/dV$ in the plane of the gate voltage
and the source-drain voltage has a diamond-shaped structure. The
behavior is  familiar to that one found in the Kondo dot, however,
it is a negative picture with dips in place of peaks. As we have
explained above, the picture results from weakening of destructive
interference processes in the system and opening the bridge
channel.

Let us now analyze electron transport though the Aharonov-Bohm
ring with the Kondo dot. In this case the hopping integrals are
complex numbers $t_{\nu}=|t_{\nu}|e^{i\phi_{\nu}}$ ($\nu=LR$,
$L0$, $R0$), where $\phi_{\nu}$ corresponds to the phase of
electronic wave passing through the $\nu$ arm of the ring in
presence of the magnetic field. At $T=0$ the conductance is given
by Eq.(\ref{12}), which simplifies the Kondo regime ($n=1$) to the
form
\begin{eqnarray}\label{13}
{\cal G}=\frac{2e^2}{h}\frac{4|t_{L0}|^2|t_{R0}|^2}{(|t_{L0}|^2
+|t_{R0}|^2)^2w^2} \bigl[1+\pi^4\rho^4|t_{LR}|^4\nonumber\\
-2\pi^2\rho^2|t_{LR}|^2 \cos(2\pi\Phi/\Phi_0)\bigr]\;.
\end{eqnarray}
Here, $\Phi=(\phi_{LR}-\phi_{L0}-\phi_{R0})/2\pi\Phi_0$  is the
magnetic flux enclosed in the ring and $\Phi_0=hc/e$ denotes the
one-electron flux quantum. For $T>0$ the electronic transport is
calculated numerically and the results are presented in Fig.3. The
conductance shows oscillations with $\Phi$ (Fig.3a) with a large
amplitude in the mixed-valence regime. In the Kondo regime the
amplitude is smaller, but strongly temperature-dependent. Figure
3b presents evolution of the voltage dependence of $dI/dV$ with
the magnetic flux $\Phi$. At $\Phi=0$ there is a dip in the curve,
which is continuously transformed to a peak for $\Phi=0.5 hc/e$.

Just recently Van der Wiel et al. \cite{wiel} have performed an
experiment on electron transport through the Aharonov-Bohm ring
made in two-dimensional electron gas with a quantum dot in one of
the arms. The zero-bias conductance, in their experiment,
increases considerably and can even reach the value of $2e^2/h$ in
some magnetic field ranges. It is in agreement with our results in
Fig.3a, where ${\cal G}$ can increase to $2e^2/h$ for the Kondo
dot if $T\to 0$ (see also \cite{gerland}). We expect that this
type of experiment \cite{yacoby,wiel} should also show
transformation from a peak to a dip in the voltage dependence of
$dI/dV$ as a function of flux through the ring (as seen in
Fig.3b).

\begin{figure}[h]\label{figfk3r}\vskip 0.2cm
\centerline{\epsfxsize=0.34\textwidth \epsfbox{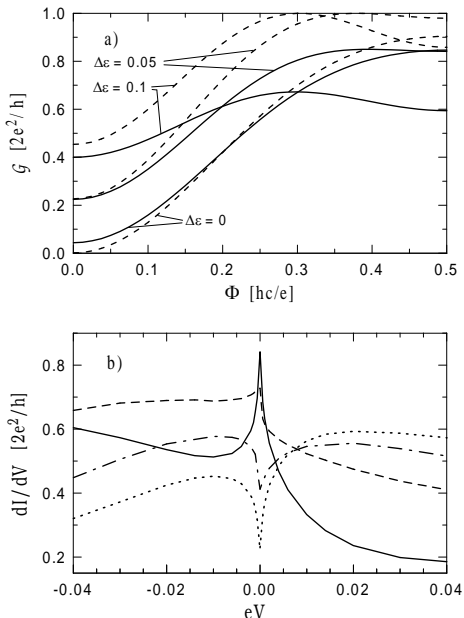}
}\vskip 0.1cm \caption{Characteristics of the Aharonov-Bohm ring
with the Kondo dot: Fig.a - the zero-bias conductance vs. the
magnetic flux for $\Delta\epsilon= 0.1$, 0.05, 0 at $T=2\times
10^{-6}$ (solid curves) and $T=0$ (dashed curves); Fig.b - the
differential conductance vs. the source-drain voltage for
$\Phi=0.5 hc/e$ (solid), $0.25 hc/e$ (dashed) , $0.125 hc/e$
(dash-dotted) and 0 (dotted) at $T=2\times 10^{-6}$,
$\Delta\epsilon= 0.05$. }
\end{figure}Summarizing, our theoretical studies of the electronic transport
through the quantum dot of the Kondo type showed that interference
of travelling waves with the localized state can lead to the Fano
effect, for which the current characteristics are strongly
modified in the Kondo regime. The source-drain voltage  dependence
of the conductance exhibits a large peak or a dip, depending on
the interference conditions. We predict that interference effects
should be pronouncedly seen in transport through the Aharonov-Bohm
ring with the Kondo dot. \vskip -0.01cm
 We have benefitted from
discussions with Gerd Czycholl and Tomasz Kostyrko. The work was
supported by the Committee for Scientific Research (KBN) under
Grant No.~2 P03B 087 19. \vskip -0.01cm {\it Note added.}$-$ After
submission of this Letter we learned of work by Heemeyer
\cite{heemeyer}, in which qualitatively similar results were
obtained for the zero-bias conductance (as in Fig.1) within the
mean-field slave-boson approach for the model with the bridge
channel.

\end{multicols}

\end{document}